\documentclass[journal]{IEEEtran}

\usepackage{enumerate}
\usepackage{graphicx}
\usepackage{epsf}
\usepackage{subfigure}
\usepackage{amsmath}
\usepackage{amssymb}
\usepackage{array}
\usepackage{setspace}
\usepackage[amsmath,thmmarks]{ntheorem}
\usepackage[ruled,lined]{algorithm2e}
\usepackage{algorithmic}
\usepackage{color}
\usepackage{amsfonts}
\usepackage{dsfont}
\usepackage{xfrac}
\usepackage{breqn}
\usepackage{bbm}
\usepackage{cite}

\graphicspath{{Fig/},{fig/}}

\theoremseparator{.}

\begin{document}

\title{\LARGE Parametrized Sharing for Multi-Agent Hybrid DRL for Multiple Multi-Functional RISs-Aided Downlink NOMA Networks}



\author{Chi-Te Kuo, Li-Hsiang Shen,~\IEEEmembership{Member,~IEEE} and Jyun-Jhe Huang}

\maketitle

\begin{abstract}
Multi-functional reconfigurable intelligent surface (MF-RIS) is conceived to address the communication efficiency thanks to its extended signal coverage from its active RIS capability and self-sustainability from energy harvesting (EH). We investigate the architecture of multi-MF-RISs to assist non-orthogonal multiple access (NOMA) downlink networks. We formulate an energy efficiency (EE) maximization problem by optimizing power allocation, transmit beamforming and MF-RIS configurations of amplitudes, phase-shifts and EH ratios, as well as the position of MF-RISs, while satisfying constraints of available power, user rate requirements, and self-sustainability property. We design a parametrized sharing scheme for multi-agent hybrid deep reinforcement learning (PMHRL), where the multi-agent proximal policy optimization (PPO) and deep-Q network (DQN) handle continuous and discrete variables, respectively. The simulation results have demonstrated that proposed PMHRL has the highest EE compared to other benchmarks, including cases without parametrized sharing, pure PPO and DQN. Moreover, the proposed multi-MF-RISs-aided downlink NOMA  achieves the highest EE compared to scenarios of no-EH/amplification, traditional RISs, and deployment without RISs/MF-RISs under different multiple access.

\end{abstract}

%

{\let\thefootnote\relax\footnotetext
{Chi-Te Kuo, Li-Hsiang Shen and Jyun-Jhe Huang are with the Department of Communication Engineering, National Central University, Taoyuan 320317, Taiwan. (email: harber9322242@gmail.com, shen@ncu.edu.tw, and kenneth900912@g.ncu.edu.tw)}}

\section{Introduction}
In the era of the six-generation (6G) wireless communications, the scarcity of spectrum resources has driven researchers to explore more efficient transmission technologies \cite{ref1}. Among them, non-orthogonal multiple access (NOMA) has emerged as a promising solution due to its capability to serve multiple users simultaneously at the same time by frequency resource \cite{ref3}. Compared to the traditional orthogonal multiple access (OMA) mechanism, NOMA exhibits significantly higher spectral efficiency. However, the performance of NOMA is often hindered by challenges such as severe channel fading and inter-user interference, potentially degrading the signal quality and limit its practical deployment \cite{ref4}. To address the issues, reconfigurable intelligent surfaces (RIS) has been proposed as an enabling technology \cite{ref5}. By adjusting the configuration of RIS elements, a virtual line-of-sight (LoS) link can be established to bypass obstacles between the transmitter and receiver and to improve the signal quality for combating the channel fading. Although RIS holds promise for next-generation systems, its practical deployment remains constrained by several inherent limitations. Particularly, it can only be operated over a 180-degree half-space coverage area, depending on external power supplies, which confines its independent operation and scalability.

Against the backdrops of RISs, the concept of multi-functional RIS (MF-RIS) has been proposed \cite{ref19, ref6}, integrating the function of simultaneous transmission and reflection RIS (STAR-RIS), providing a 360-degree full-space coverage, realizing a ubiquitous service \cite{ref8, my1}. The work of \cite{ref19} has established the fundamental system modeling and corresponding performance optimization of the MF-RIS deployment. Moreover, energy harvesting (EH) capability at the radio-frequency (RF) is designed in MF-RIS, allowing it to capture wireless energy from incident electromagnetic signals and operating in a self-sustainable manner \cite{new2, ref13}. This design reduces the requirements on the wired power infrastructure or frequent battery replacement, improving system energy efficiency (EE) and deployment flexibility \cite{my2}. The authors in \cite{new2} also conduct sum-rate maximization with robust and secure transmissions by MF-RIS. The work of \cite{new1} has analyzed the feasibility and its performance when leveraging MF-RIS with self-sustainable capability. Additionally, MF-RIS enhances the traditional passive reflection by incorporating active components for signal amplification, improving the weak channel conditions in NOMA networks \cite{ref11, ref12, ref15}. Moreover, there will be increasing needs of deploying multiple MF-RISs for wider coverage requirements. Extending from prior work of \cite{ref19} with signal MF-RIS and rate optimization, we investigate a novel architecture of deploying multiple MF-RISs for assisting downlink NOMA networks \cite{ref16}, associated with the EE optimization. NOMA shares the same spectrum in multi-MF-RISs-aided networks. On the other hand, MF-RISs contribute to constructing favorable channel conditions for NOMA user groups by mitigating channel fading and interference effects. Moreover, we design based on deep reinforcement learning (DRL) \cite{ref2} techniques to enable adaptive policy learning under high-dimension and dynamic environments. Unlike conventional DRL handling either discrete or continuous actions separately, a general hybrid DRL framework should be adopted to effectively address complex hybrid continuous-discrete action spaces. 

The main contributions of this work are summarized as follows: 
\textbf{(1)} 
We investigate multi-MF-RISs-aided downlink NOMA networks. We consider power-domain NOMA, where a group of users shares the same frequency resource. MF-RISs are capable of extending the transmission range by reflecting, transmitting, and amplifying signals, while harvesting partial signal energy for operation. 
\textbf{(2)} We aim at maximizing system EE by deciding power allocation, base station (BS) beamforming and MF-RIS configurations of amplification/phase-shifts/EH ratios and MF-RIS positions. Note that MF-RIS circuit power is also considered. A parametrized sharing in multi-agent hybrid deep reinforcement learning (PMHRL) scheme is designed, whereas hybrid DRL tackles joint continuous-discrete variables respectively by proximal policy optimization (PPO) and deep-Q network (DQN). Parametrized sharing enables information sharing between dual-modules.
\textbf{(3)} Results have demonstrated that PMHRL achieves the highest EE compared to other existing benchmarks of conventional DRLs and those without parametrized sharing. The proposed architecture of multi-MF-RISs-aided downlink NOMA achieves the highest EE among the cases without EH, conventional RISs, non-amplified signals and deployments without RISs/MF-RISs.
%
%
%

\section{System Model and Problem Formulation}

\begin{figure}[!t]
\centering
\includegraphics[width=2.5in]{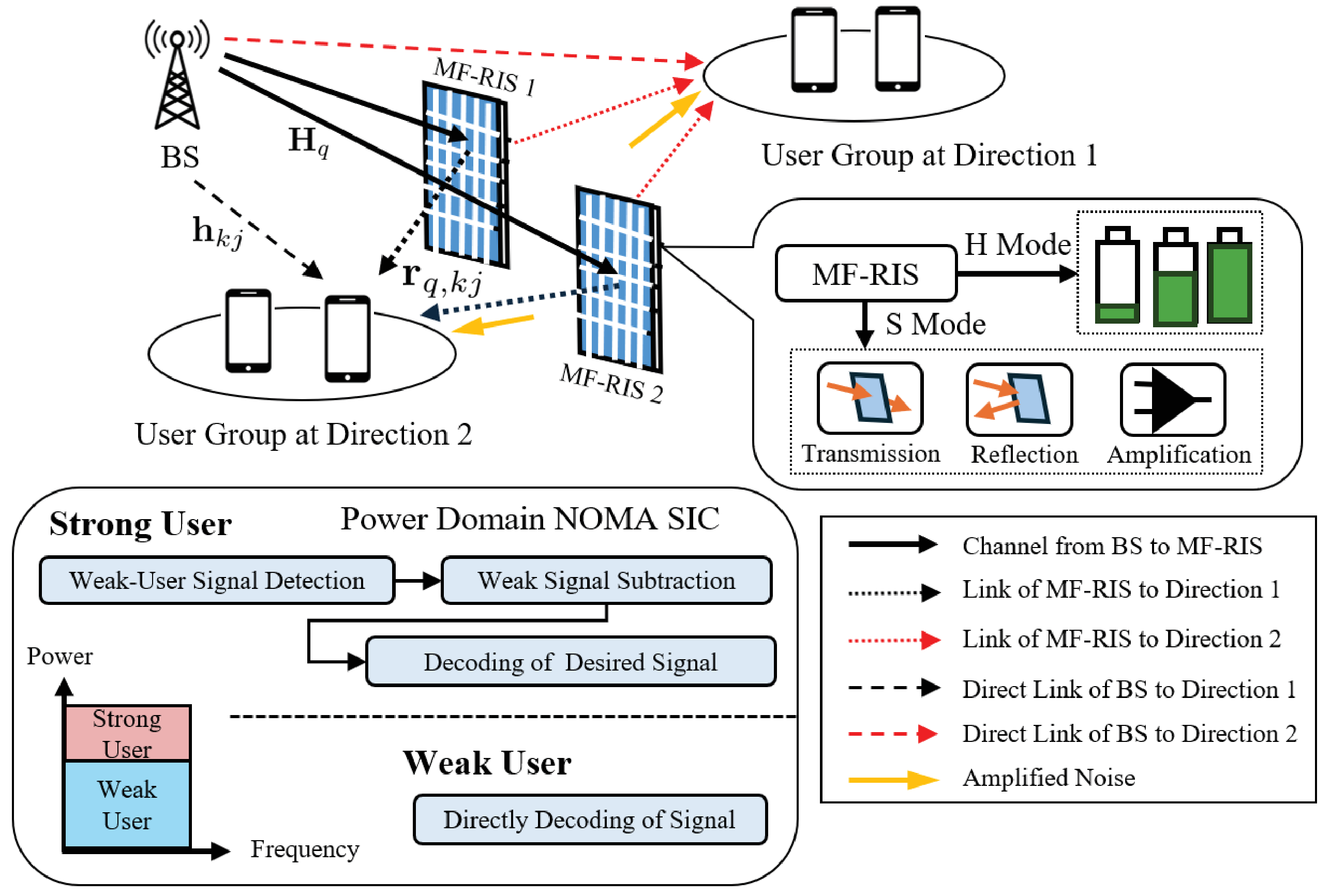}
\caption{The proposed architecture of multi-MF-RISs-assisted downlink NOMA.} \label{Fig.1}
\end{figure}

	In a multi-MF-RISs-assisted downlink NOMA network in Fig. \ref{Fig.1}, we consider a BS equipped with $N$ transmit antennas with the set of $\mathcal{N} = \{1, 2, . . . , N\}$, serving $J$ users at direction $k$ indexed by the set of $\mathcal{J}_k = \{1, 2, . . . , J_k\}$. We consider total $K$ direction for NOMA transmission groups, where $\mathcal{K}=\{1,2.....K\}$. We consider $Q$ MF-RISs with its set of $\mathcal{Q} = \{1, 2, \ldots, Q\}$. Furthermore, we consider a Cartesian coordinate system with the locations of the BS, MF-RIS, and user being $\mathbf{w}_b = [x_b, y_b, z_b]^{\mathrm{T}}$, $\mathbf{w}_q = [x_q, y_q, z_q]^{\mathrm{T}}$, and $\mathbf{w}_{kj} = [x_{kj}, y_{kj}, 0]^{\mathrm{T}}$, respectively. Note that $\mathrm{T}$ indicates the transpose operation. Due to the limited coverage of MF-RIS, its deployable region is also limited by $\mathcal{W}$, where the following constraint is satisfied: $\mathbf{w}_q \in \mathcal{W} = \{ [x_q, y_q, z_q]^{\mathrm{T}} | \mathbf{w}_{\min} \preceq \mathbf{w}_{q} \preceq \mathbf{w}_{\max} \}$ with its deployable areas bounded by $\mathbf{w}_{\min}$ and $\mathbf{w}_{\max}$. Each MF-RIS has $M$ elements indexed by the set of $\mathcal{M} = \{1, 2, \ldots, M\}$ with a two-dimensional array with $M = M_h  \cdot M_v$ elements, where $M_h$ and $M_v$ indicate the respective numbers of elements in horizontal and vertical axes. Each MF-RIS configuration can be defined as $\mathbf{\Theta}_{q}^k = \operatorname{diag}\left( 
    \alpha_{q,1} \sqrt{\beta_{q,1}^k} e^{j\theta_{q,1}^k}, \ldots, \alpha_{q,M} \sqrt{\beta_{q,M}^k} e^{j\theta_{q,M}^k} 
\right)$,
where $ \theta_{q,m}^k \in [0, 2\pi) $ and $ \beta_{q,m}^k \in [0, \beta^k_{\max}] $ denote the phase-shift and amplitude coefficients of MF-RIS at $k$-th direction, respectively. Note that $\beta_{\max} > 1$ denotes the signal amplification, whereas $\beta_{\max}  \leq 1$ indicates conventional RIS without amplification capability. Each element of the MF-RIS can operate in harvesting mode (H mode) and signal mode (S mode) by adjusting the EH coefficient $\alpha_{q,m}\in \{0, 1\}$. Note that $\alpha_{q,m} = 1$ implies that MF-RIS operates in S mode, whilst $\alpha_{q,m} = 0$ indicates that it functions in only H mode.


We consider the Rician fading channel model between the BS and $q$-th MF-RIS as $\mathbf{H}_{q} = \sqrt{h_0 d_{q}^{-k_0}}
\left( \sqrt{\frac{\beta_0}{\beta_0 + 1}} \mathbf{H}_q^{\text{LoS}} + \sqrt{\frac{1}{\beta_0 + 1}} \mathbf{H}_q^{\text{NLoS}} \right) \in \mathbb{C}^{M \times N}$,
where $h_0$ is the pathloss at the reference distance of 1 meter, $d_{q}= \|\mathbf{w}_b-\mathbf{w}_q\|^2 $ is the distance, and $k_0$ is the pathloss exponent. $\beta_0$ is the Rician factor adjusting the portion of LoS path $\mathbf{H}_q^{\text{LoS}}$ and non-LoS (NLoS) component of $\mathbf{H}_q^{\text{NLoS}}$. The LoS component \cite{ref19} is expressed as 
$
\mathbf{H}_q^{\text{LoS}} = 
\begin{bmatrix}
1,e^{-j\frac{2\pi}{\lambda}d_R\sin\bar{\psi}_{r,q}\sin\bar{\theta}_{r,q}}, \cdots
,e^{-j\frac{2\pi}{\lambda}(M_z\!-\!1) d_R\sin\bar{\psi}_{r,q}
\sin\bar{\theta}_{r,q}}
\end{bmatrix}^{\mathrm{T}} 
	\otimes
\begin{bmatrix}
1,e^{-j\frac{2\pi}{\lambda}d_R\sin\bar{\psi}_{r,q}\cos\bar{\theta}_{r,q}}, \cdots,e^{-j\frac{2\pi}{\lambda}(M_y \!-\! 1)d_R\sin\bar{\psi}_{r,q}\cos\bar{\theta}_{r,q}}
\end{bmatrix}^{\mathrm{T}} 
	\! \otimes\!
\begin{bmatrix}
1,e^{-j\frac{2\pi}{\lambda}d_B\sin\varphi_t\cos\vartheta_t}, \cdots,e^{-j\frac{2\pi}{\lambda}(N\!-\!1)d_B\sin\varphi_t\cos\vartheta_t}
\end{bmatrix}^{\mathrm{T}}
$,
where $\otimes$ denotes the Kronecker product and $T$ is transpose operation. $\lambda$ indicates the wavelength of the operating frequency. Notations of $d_R$ and $d_B$ denote the element spacing of MF-RIS and antenna separation of BS, respectively. Notations of $\bar{\psi}_{r,q}$, $\bar{\theta}_{r,q}$, $\varphi_t$, and $\vartheta_t$ represent the azimuth and elevation angles of-arrivals of MF-RIS $q$, and those of angle-of-departures of BS, respectively. Note that $\mathbf{H}^{\text{NLoS}}_q$ follows the Rayleigh fading. The direct link from BS and reflected link from the MF-RIS $q$ to user $j$ at direction $k$ are denoted by $\mathbf{h}_{kj} \in \mathbb{C}^{N \times1 }$ and $\mathbf{r}_{q,kj} \in \mathbb{C}^{M \times 1 }$, respectively, associated with their distances of $d_{kj}$ and $d_{q,kj}$. While, both parameters follow $\mathbf{H}_{q} $ but in a vector form, where the LoS components are $\mathbf{h}^{\text{LoS}}_{kj} \!=\! [
1, e^{-j\frac{2\pi}{\lambda}d_B\sin\varphi_t\sin\vartheta_t}, \cdots, e^{-j\frac{2\pi}{\lambda}(N\!-\!1)d_B\sin\varphi_t\sin\vartheta_t}
]^{\mathrm{T}}$ and $\mathbf{r}_{q,kj}^{\text{LoS}} = [
1, e^{-j\!\frac{2\pi}{\lambda}\!d_R\sin\!\varphi_{t,q}\sin\!\vartheta_{t,q}}, \cdots, e^{-j\!\frac{2\pi}{\lambda}\!(M\!-\!1)d_R\sin\!\varphi_{t,q}\sin\!\vartheta_{t,q}}
]^{\mathrm{T}}$. The NLoS parts $\mathbf{h}_{kj}^{\text{NLoS}}$ and $\mathbf{r}_{q,kj}^{\text{NLoS}}$ are both characterized by Rayleigh fading. Accordingly, the channel between the MF-RIS $q$ and and user 
$j$ at direction $k$ is $\mathbf{g}_{q,kj} = \mathbf{r}_{q,kj}^H \mathbf{\Theta}_{q}^k \mathbf{H}_{q}$ where $H$ indicates Hermitian operation. The total combined channel of BS-user $j$ at direction $k$ assisted by $Q$ MF-RISs is defined as
${\mathbf{g}}_{kj} = \mathbf{h}_{kj} + \sum_{q\in\mathcal{Q}} \mathbf{g}_{q,kj}$.

In the downlink-NOMA network, users are divided into multiple groups to share spectrum resources. The signal received of user $j$ at direction $k$ is given by
\begingroup
\allowdisplaybreaks
\begin{align} \label{r_signal}
& y_{kj} = \mathbf{g}_{kj} \mathbf{f}_k \sqrt{p_{kj}} s_{kj}+\mathbf{g}_{kj} \mathbf{f}_k \sum_{i \in \mathcal{J}_k \setminus \{j\}} \sqrt{p_{ki}} s_{ki} \notag \\
& +\sum_{\bar{k} \in \mathcal{K} \setminus \{k\} }\mathbf{g}_{kj} \mathbf{f}_{\bar{k}}
\sum_{i \in \mathcal{J}_{\bar{k}}}\sqrt{p_{\bar{k}i}} s_{\bar{k}i} + \sum_{q \in \mathcal{Q}} \mathbf{r}^H_{q,kj}\mathbf{\Theta}_{q}^k \mathbf{n}_q + {n}_{kj},       
\end{align}
\endgroup
where $\mathbf{f}_k$ represents the transmit beamforming vector of the BS for direction 
$k$. Moreover, $p_{kj}$ denotes the power allocation factor for user $j$ at direction $k$ where $\sum_{j \in \mathcal{J}_k} p_{kj} = 1$. $\mathbf{n}_{q} \sim \mathcal{CN}(0, \sigma_s^2 \mathbf{I}_M)$ denotes the amplification noise from MF-RISs with element noise power $\sigma_s^2$. Notation of $n_{kj}$ is noise power of user $j$ at direction $k$ with its power of $\sigma_u^2$. Moreover, NOMA user signals are transmitted simultaneously at the same frequency, leading to mutual interference. To decode the intended signals, users employ successively interference cancellation (SIC) \cite{ref14}. Assume that the users $j$ and $l$ in direction $k$ are ranked in an ascending order according to the equivalent combined channel gains, i.e., $|{\mathbf{g}}_{k1}|^2 \geq ... \geq |{\mathbf{g}}_{kj}|^2 \geq |{\mathbf{g}}_{kJ}|^2$, associated with the SIC conditions of
\begingroup
\allowdisplaybreaks
\begin{align} 
 \frac{|{\mathbf{g}}_{kj}^H \mathbf{f}_k|^2}{
|{\mathbf{g}}_{kl}^H \mathbf{f}_k|^2 + I_{kj} 
+ \sigma_u^2 } \leq 
\frac{|{\mathbf{g}}_{kl}^H \mathbf{f}_k|^2}{
|{\mathbf{g}}_{kj}^H \mathbf{f}_k|^2 
+ I_{kl} 
+ \sigma_u^2 }, \label{9}
\end{align}
\endgroup
where $k \in \mathcal{K}$, $j \in \mathcal{J}_k$ denote users at direction $k$, and $l \in \mathcal{L}_k = \{j, j+1, \cdots, J_k\}$. Notation $I_{kj} = \sum_{q \in \mathcal{Q}} \sigma_s^2 \| \mathbf{r}_{q,kj}^H \mathbf{\Theta}_q^k \|^2$ indicates the residual interference. Note that inter-/intra-group interferences are considered constants and is negligible in SIC conditions in \eqref{9}. However, such effects should be taken into account in calculating signal-to-interference-plus-noise ratio (SINR) given by $\gamma_{kj}=\frac{| \mathbf{g}_{kl} \mathbf{f}_k|^2 p_{kj}}{
\sum_{l \in \mathcal{L}_k} |\mathbf{g}_{kl} \mathbf{f}_k|^2 p_{kl} + I_{\text{IG},k} + I_{kj} + \sigma_u^2}$, 
where $I_{\text{IG},k} = \sum_{\bar{k} \in \mathcal{K} \setminus \{k\} } \sum_{i \in \mathcal{J}_{\bar{k}}} \| \mathbf{g}_{kj} \mathbf{f}_{\bar{k}} \|^2 p_{\bar{k}i}$ indicates the inter-group interference. Note that the first term of the denominator of SINR means the intra-group interference after SIC. Therefore, the achievable rate for user $j$ at direction $k$ can be expressed as $R_{kj} = \log_2(  1 + \gamma_{kj})$.

    Here, we define the EH coefficient matrix for the $m$-th element of the $q$-th MF-RIS as $\mathbf{T}_{q,m}=\operatorname{diag} \left( \left[ {0, \dots, 0}, 
1 - \alpha_{q,m}, {0, \dots, 0}\right] \right)$. Therefore, the RF power recieved by the $m$-th element of the $q$-th MF-RIS is given by $P_{q,m}^{\text{RF}}=\mathbb{E} \left( \left\| \mathbf{T}_{q,m} \left( \mathbf{H}_{q} \sum_{k \in \mathcal{K}}\mathbf{f}_k + \mathbf{n}_{q,m} \right) \right\|^2 \right)$, where $\mathbf{n}_{q,m}$ is the amplified noise introduced by the MF-RIS. To capture RF energy conversion efficiency for different input power, a non-linear harvesting model is adopted. Accordingly, the total power of the $m$-th element of the $q$-th MF-RIS is expressed as $P_{q,m}^{\text{A}} = \frac{\Upsilon_{q,m} - Z\Omega}{1 - \Omega}$, where $\Upsilon_{q,m}=\frac{Z}{1 + e^{-p \left( P_{q,m}^{\text{RF}} - k \right)}}$ is a logistic function with respect to the received RF power $P^{\text{RF}}_{q,m}$, and $Z>0$ is a constant determining the maximum harvested power. Constant $\Omega=\frac{1}{1+e^{\varpi_1 \varpi_2}}$ ensures a zero-input/zero-output response in H mode with constants $\varpi_1>0$ and $\varpi_2 > 0$ capturing the effects of circuit sensitivity and current leakage. Moreover, the power for controlling MF-RIS mainly comes from the total number of PIN diodes required \cite{ref17-1}. The quantization levels assigned for EH ratio, amplitude and phase shifts are $L_\alpha$, $L_\beta$, and $L_\theta$, respectively, where the total number of PIN diodes per MF-RIS is $\log_2 L_{\alpha} + K \log_2 L_{\beta} + K \log_2 L_{\theta}$. We have the following self-sustainability constraint per MF-RIS, i.e., $P_{q}^{\text{con}} \leq \sum_{m \in \mathcal{M}} P_{q,m}^{\text{A}}$, where $P_{q}^{\text{con}}= \lceil \log_{2} L_{\alpha} + K \log_{2} L_{\beta} + K \log_{2} L_{\theta} \rceil \cdot M P_{\text{PIN}}
+ P_{C} + \xi \cdot P_{l,O}$. Here, $P_C$ denotes the power consumption of RF-to-DC power conversion, and $P_{\text{PIN}}$ is power consumption per PIN diode. Notation $\xi$ indicates the inverse of amplifier efficiency. The output power of MF-RIS $q$ is obtained as $ P_{O,q} = \sum_{k \in \mathcal{K}} \left( \sum_{k' \in \mathcal{K}} \|\boldsymbol{\Theta}_q^k \mathbf{H}_q \mathbf{f}_{k'}\|^2 + \sigma_s^2 \|\boldsymbol{\Theta}_q^k\|_F^2 \right)$, where $\| \cdot \|_F$ is Forbenius norm.

The objective is to maximize the system EE while guaranteeing the constraints of minimum user rate requirement, MF-RIS configuration and power limitation, which is formulated as
\begingroup
\allowdisplaybreaks
\begin{subequations} \label{prob}
    \begin{align}
    \max_{\substack{p_{kj}, \mathbf{f}_{k}, \alpha_{q,m}, \\ \beta_{q,m}^k, \theta_{q,m}^k, \mathbf{w}_q}} 
    &\quad \sum_{k \in \mathcal{K}} \sum_{j \in \mathcal{J}_k} \frac{R_{kj}}{P_{\text{total}}} \label{prob_25a} \\
    \text{s.t. } \quad
    & \eqref{9}, \ \mathbf{\Theta}_q \in \mathcal{R}_{\mathbf{\Theta}}, \ \mathbf{w}_q \in \mathcal{W}, \quad \forall q \in \mathcal{Q}, \label{prob_25b} \\
    &R_{kj} \geq R^{\min}_{kj}, \quad\qquad \forall k \in \mathcal{K}, \, \forall j \in \mathcal{J}_k, \label{prob_25c} \\
    &\sum_{j \in \mathcal{J}_k} p_{kj} = 1, \quad\qquad \forall k \in \mathcal{K}, \label{prob_25d} \\
    &\sum_{k \in \mathcal{K}} \left\lVert \mathbf{f}_k \right\rVert^2 \leq P_{BS}^{\max}, \label{prob_25e} \\
    &P_q^{\text{con}} \leq \sum_{m \in \mathcal{M}} P_{q,m}^{\text{A}}, \quad \forall q \in \mathcal{Q}, \label{prob_25f} 
    \end{align}
\end{subequations}
\endgroup
where $P_{\text{total}}=\sum_{q \in \mathcal{Q}}(P_q^{\text{con}} -\sum_{m \in \mathcal{M}} P_{q,m}^{\text{A}})+\sum_{k \in \mathcal{K}} \left\lVert \mathbf{f}_k \right\rVert^2$ is the system total consumed power. Constraint set in $\mathcal{R}_{\mathbf{\Theta}}$ in \eqref{prob_25b} specifies the feasible region of MF-RISs, i.e., $\alpha_{q,m} \in [0,1]$, $\beta_{q,m}^k \in [0, \beta_{\max}^k]$, $\theta_{q,m}^k \in [0, 2\pi)$. \eqref{prob_25c} ensures the minimum rate requirement per user as $R^{\min}_{kj}$, while \eqref{prob_25d} represents the NOMA power allocation restriction. Constraint \eqref{prob_25e} ensures that the total BS transmit power cannot exceed its budget $P_{BS}^{\max}$. \eqref{prob_25f} guarantees MF-RIS self-sustainability. Due to non-convexity and non-linearity of problem \eqref{prob}, we propose a DRL-based scheme as detailed in the following section.

\section{Proposed PMHRL Scheme}



\subsection{Hybrid DRL Algorithm}
	We consider a multi-agent hybrid DRL framework characterized by state space $\mathcal{S}$, action space $\mathcal{A}$, and reward $\mathcal{R}$. Within this framework, each agent corresponds to a single MF-RIS, which interacts with the dynamic environment by taking actions, receiving rewards, and updating its local states accordingly. In addition, the BS is also considered as an independent agent, responsible for controlling power allocation and transmit beamforming. Conventional DRL methods struggle under conditions of high complexity, slow convergence and instability during training. Moreover, the use of pure DQN or PPO becomes compellingly impractical, as both quantizing continuous variables into discrete ones and recovering continuous parameters from quantized ones introduce extra computational overhead and potential quantization errors. To overcome these challenges, we adopt a hybrid DRL architecture that incorporates both DQN and PPO networks for efficiently handling discrete and continuous action spaces separately. We define the state, action, and the corresponding reward as follows:
 \begin{itemize}
    \item \textbf{State:} The total state space is defined as a set of individual agent state $\mathcal{S}(t) = \{s_1(t), s_2(t), \ldots, s_Q(t), s_{Q+1}(t)\}$. Each agent state $s_q(t)$ is designed as $s_q(t) = \{\mathbf{g}_{q,kj}(t) | \forall k \in \mathcal{K}, \forall j \in\mathcal{J}_k\}, \forall q\in\mathcal{Q}$ and $s_{Q+1}(t) = \{\mathbf{g}_{kj}(t) | \forall k \in \mathcal{K}, \forall j \in\mathcal{J}_k \}$ associated with the combined channel at timestep $t$. Note that index $1\leq q\leq Q$ indicates the MF-RIS agents, whereas $q=Q+1$ stands for the BS agent.

    \item \textbf{Action:} The action space is defined as a set of individual action $ \mathcal{A}(t) = \{ a_1(t), a_2(t), \ldots, a_Q(t), a_{Q+1}(t)\}\ $. For agents representing MF-RIS $q \in \mathcal{Q}$, each action $a_q(t)= \{ a_q^{\text{dis}}(t), a_q^{\text{con}}(t)\}$ is composed of both discrete and continuous components. Specifically, the discrete action corresponds to the selection of mode of each MF-RIS element, defined as $a_{q}^{\text{dis}}(t) = \{\alpha_{q,m}|\forall m \in \mathcal{M}\}$. On the other hand, the continuous action includes MF-RIS configurations of amplitude and phase-shifts as well as MF-RIS position, denoted as  $a^{\text{con}}_q(t) = \{ \beta^{k}_{q,m}, \theta^{k}_{q,m}, \mathbf{w}_q | \forall k \in \mathcal{K},  \forall m \in \mathcal{M}\}$. In addition, For the $(Q+1)$-th agent representing the BS, the output action consists of only continuous variables, i.e., the power allocation for NOMA users and the beamforming vectors, defined as $a_{Q+1}(t) = \{ p_{kj},\mathbf{f}_k | \forall k\in\mathcal{K}, j\in\mathcal{J}_k\}$.
    
    \item \textbf{Reward:} 
We design the shared reward as the overall EE in conjunction with its constraints as penalties, given by $r(t) = \frac{\sum_{k\in \mathcal{K}}\sum_{j\in \mathcal{J}_k} R_{kj}}{P_{\text{total}}} - \sum_{i=1}^{3} \rho_{i} C_{i}$,
where $\rho_{i}, \forall i\in \{1,2,3\}$ indicates the weights of each penalty $C_{i}$ corresponding to constraints of \eqref{prob_25c}, \eqref{prob_25e}, and \eqref{prob_25f}, which are defined as $ C_{1} = \sum_{k\in\mathcal{K}} \sum_{j\in\mathcal{J}_k} (R_{kj}^{\min} - R_{kj})$, $C_{2} = \sum_{k \in \mathcal{K}}\mathbf \|\mathbf{f}_k\|^2 - P_{BS}^{\max}$, and $ C_{3} = \sum_{q\in \mathcal{Q}}( P_{q}^{\text{con}} -\sum_{m \in \mathcal{M}} P_{q,m}^{\text{A}})$, respectively. Note that the remaining boundary conditions in \eqref{prob_25b} and \eqref{prob_25d} can be automatically constrained during generating actions.
\end{itemize}

\subsubsection{DQN for Discrete Variables}

DQN employs a deep neural network, Q-network, to approximate the Q-function  $Q_q(s_q, a_q| \omega_q^{\phi})$ which estimates the expected cumulative reward for each action $a_q$ under a given state $s_q$. Note that we define $\omega^{\phi}_q$ and $\omega^{\phi^{-}}_q$ as the model weights of the current network and of the target network of DQN, respectively. Based on estimated Q-values, each agent selects its action using an $\epsilon$-greedy strategy, which balances exploration and exploitation by choosing a random action with probability $\epsilon$ and by selecting the action with the maximum predicted Q-value with probability $ 1 - \epsilon $, i.e., $\epsilon(t) = \epsilon({t-1})-\frac{\epsilon_{\max} - \epsilon_{\min}}{\epsilon_d}$, where $\epsilon_d$ is the decay parameter. Notations of $\epsilon_{\max}$ and $\epsilon_{\min}$ indicate the maximum and minimum exploration boundaries, respectively. To enhance training stability, DQN incorporates two critical techniques: (i)  \textit{Experience replay buffer} stores historical trajectories with a tuple of $ (s_q, a_q, r_q, s'_q)$, allowing the agent to sample mini-batches uniformly and eliminate temporal correlations during learning, where $s'_q$ indicates the new state; and (ii) \textit{Target network}, with its model denoted as $Q'_q(s_q, a_q| \omega_q^{\phi^-})$ is periodically softly updated to provide stable Q-learning, i.e., $\omega_q^{\phi^-} \leftarrow \tau_{\phi}\omega_q^{\phi}+(1-\tau_{\phi})\omega_q^{\phi^-}$ where $\tau_q^{\phi}$ indicates the importance of target model of DQN. The Q-network is then updated by minimizing the temporal-difference (TD) error, which measures the discrepancy between the predicted Q-value and the target Q-value, given by $\mathcal{L}(\omega_{q}^\phi) = 
\mathbb{E}_{(s_{q},a_{q},r,s'_{q})}
[
(
y - Q_{q} (s_{q}, a_{q} \mid \omega_{q}^\phi)
)^{2}
],$
where $y = r(t) + \gamma^{d} \max_{a'} Q'_{q} (s'_{q}, a' \mid \omega_{q}^{\phi^-})$ indicates the TD target value and the discount factor $ \gamma^d \in [0, 1] $ indicates the importance of future rewards. Notation of $Q'_q(\cdot)$ is the Q-value of the target network. The parameter update via gradient descent is then given by $\omega_{q}^\phi \leftarrow \omega_{q}^\phi - l^{d} \cdot \nabla_{\omega_{q}^\phi} \, \mathcal{L}(\omega_{q}^\phi)$, where $l^d \in [0, 1] $ is the learning rate.

\subsubsection{PPO for Continuous Variables}

The remaining continuous parameters are optimized using the PPO algorithm \cite{ref21}. Particularly, PPO adopts an actor-critic framework respectively consisting of a policy network and of a value network. In the policy network, the neural network outputs the mean and standard deviation of a multivariate Gaussian distribution, from which actions are sampled according to the current state $s_q(t)$ and policy  $\pi_{\delta_q}(a_q(t)|s_q(t))$. Note that $\delta_q$ indicates the policy network parameters. To optimize the policy network, we employ a clipped surrogate objective function expressed as
$\mathcal{L}^{\text{clip}} (\delta_q) 
= {\mathbb{E}}_t \big[ \min\big( O_q(\delta_q)\hat{A}_q(t),  \mathrm{clip}( O_q(\delta_q), 1 - \Lambda, 1 + \Lambda)\hat{A}_q(t) \big) \big],$
where ${\mathbb{E}}[\cdot]$ is the expectation over a batch of generated trajectories, and $O_q(\delta_q) = \frac{\pi_{\delta_q}(a_q(t)|s_q(t))}{\pi_{\delta_{q,\text{old}}}(a_q(t)|s_q(t))}$ is the probability ratio. $\mathrm{clip}(\cdot)$ indicates the clipping function which clips the change between the new and old policies within the range $[1-\Lambda, 1+\Lambda]$ for avoiding excessive policy updates. Note that $\delta_{q,\text{old}}(\cdot)$ is the old policy parameters. Furthermore, $\hat{A}_q(t)$ is the generalized advantage estimation (GAE) quantifying the difference between the observed outcome of each action in a state and the predicted state value $ V_{\mu_q}(s_q(t))$ by the value network, which is $\hat{A}_q(t) \!=\! \sum_{i=t}^{T-t}(\gamma^p\lambda^p)^{i-t} \{ [ (r(i) \!+\! \gamma^p V_{\mu_q} (s_q(i\!+\!1)) ] \!-\! V_{\mu_q} (s_q(i)) \}$,
where $\gamma^p$ and $\lambda^p$ are importance ratio and GAE hyperparameters, respectively. Notation $T$ means the length of trajectory segment. The policy is optimized iteratively by maximizing the clipped surrogate objective using gradient ascent given by $\delta_q \leftarrow {\delta_q} - l^p_{\text{ac}} \cdot \nabla_{\delta_q} \mathcal{L}^{\text{clip}} (\delta_q)$, where $l^p_{\text{ac}} \in [0, 1] $ is learning rate for actor in PPO. The associated loss function of the value network is defined as $\mathcal{L}^{V}(\mu_q) = \mathbb{E}_t [  \big( V_{\mu_q}(s_q(t)) - \hat{V}^{\text{tar}}_q(t) \big)^2]$, where $\hat{V}^{\text{tar}}_q(t) 
= V_{\mu_q^-}\!\big(s_q(t)\big) + \hat{A}_q(t)$ and $\mu_q^-$ indicates the previous update of value network. The parameters of value network are updated by the gradient method, i.e., $\mu_q \leftarrow {\mu_q} - l^p_{\text{cr}} \cdot \nabla_{\mu_q} \mathcal{L}^{V} (\mu_q)$, where $l^p_{\text{cr}} \in [0, 1] $ is learning rate for critic in PPO. Note that coordination is implicitly achieved through shared environment feedback. MF-RIS position adjustments modify the wireless channel state, which is reflected in the observed channel conditions and reward feedbacks from the BS/MF-RIS. As the learning process operates on short decision intervals, the BS/MF-RIS can quickly adapt their strategies based on the updated channels without requiring explicit inter-agent communications.

\subsection{Parametrized Sharing in PMHRL}

Individual agent design in hybrid DRL, PPO and DQN select actions independently based on their own input states. This isolated decision-making process neglects the potential interdependence and interaction between two strategies. To address this, inspired by \cite{ref23}, we propose a parametrized sharing mechanism to enable shared representation between PPO and DQN by exchanging features. Since PPO handling high-dimensional actions performs a more complex task than DQN, information from DQN is essential to PPO. Specifically, PPO agent utilizes the discrete action output from the DQN agent as the input of PPO, i.e., $s_q^{\text{con}}(t) = \operatorname{concat}(\mathbf{g}_{q,kj}(t), \operatorname{vec}(a_q^{\text{dis}}(t-1)))$, where ${\rm concat}(\cdot)$ indicates the concatenation of two vectors and ${\rm vec}(\cdot)$ vectorizes the discrete action. Note that only MF-RISs require parametrized sharing as they have hybrid actions, thereby improving coordination between the two decision modules.

\section{Simulation Results}

\begin{table}[!t]
\scriptsize
\caption{Simulation Parameters} \label{param}
\centering
\begin{tabular}{|p{2.1cm}|p{5cm}|}
\hline
\textbf{Parameter} & \textbf{Value} \\ \hline
Communication parameters & 
$h_0 = -20$ dB, $k_0 = 2.2$, $\beta_0  = 3$ dB, $\sigma_{s}^2 = \sigma_u^2 = -70$ dBm \\ \hline
MF-RIS power consumption
parameters \cite{ref19, ref17-1} &
$\xi = 1.1$, $P_{\text{PIN}} = 0.33$ mW, $P_{C} = 2.1$ mW, $Z = 24$ mW,  
$\varpi_1= 150$, $\varpi_2 = 0.014$, $L_{\alpha}=2$, $L_{\beta}=10$, $L_{\theta}=8$\\ \hline
Other parameters &  
$P_{BS}^{\max}=40$ dBm, $\mathbf{w}_{\min}=[5,10,10]$ m, $\mathbf{w}_{\max}=[5,45,10]$ m 
 \\ \hline
\end{tabular}
\end{table}

\begin{table}[!t]
\scriptsize
\caption{EH Efficiency} \label{EHcoeff}
\centering
\begin{tabular}{|c|c|c|c|c|}
\hline
$(\varpi_1,\varpi_2)$ & $(100,0.014)$ & $(150,0.014)$ & $(200,0.014)$ & $(300,0.014)$ \\  \hline
Efficiency & $0.49$ & $0.42$ & $0.28$ & $0.13$
\\  \hline\hline
$(\varpi_1,\varpi_2)$ & $(150,0.010)$ & - & $(150,0.018)$ & $(150,0.022)$ \\  \hline
Efficiency & $0.69$ & -& $0.24$ & $0.13$\\ \hline 
\end{tabular}
\end{table}

\begin{figure}[!t]
	\centering
	\subfigure[]{\includegraphics[width=1.62 in]{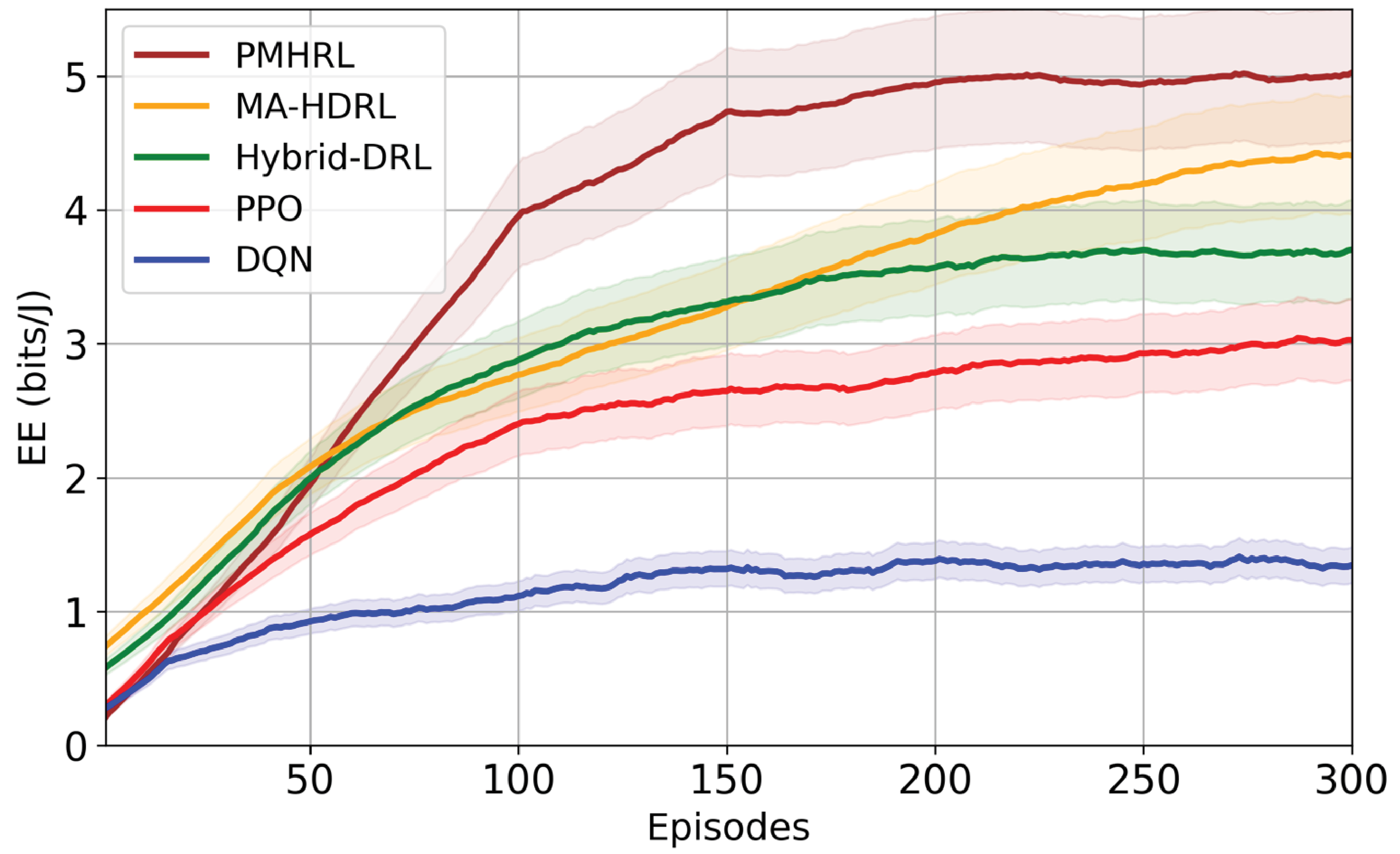} \label{Fig.4}}
	\subfigure[]{\includegraphics[width=1.62 in]{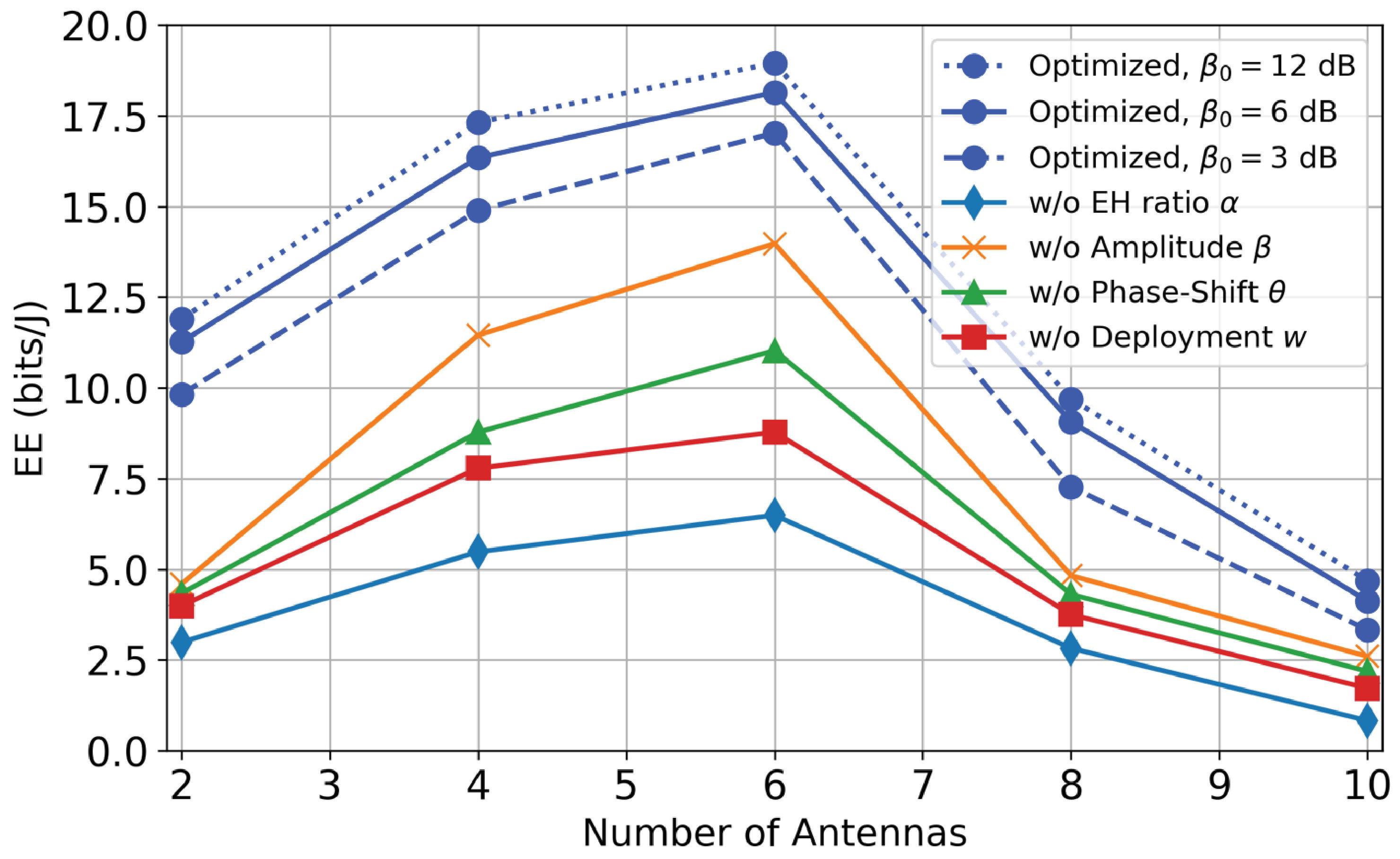} \label{Fig.4-1}}
	\subfigure[]{\includegraphics[width=1.62 in]{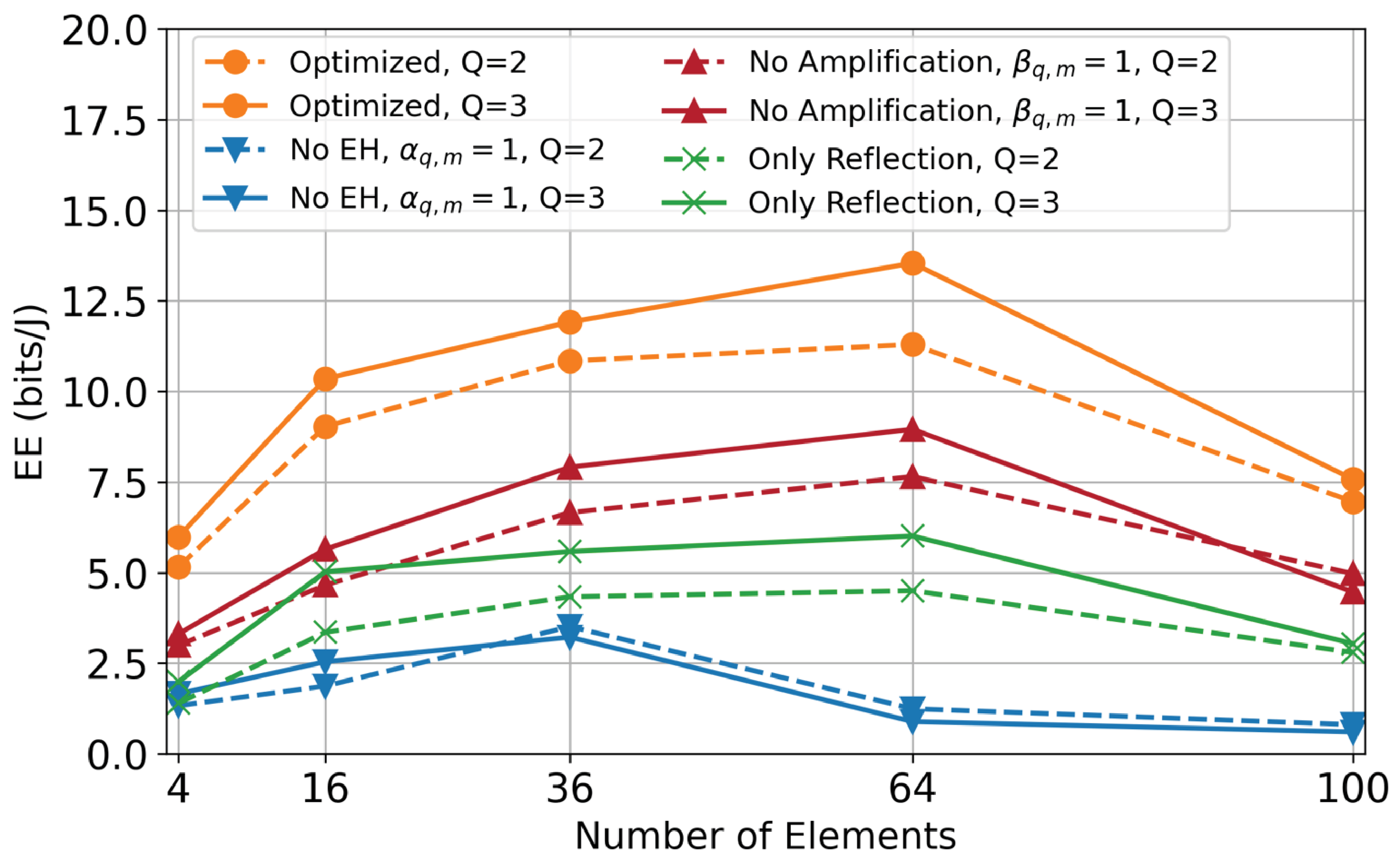} \label{Fig.5}}
	\subfigure[]{\includegraphics[width=1.62 in]{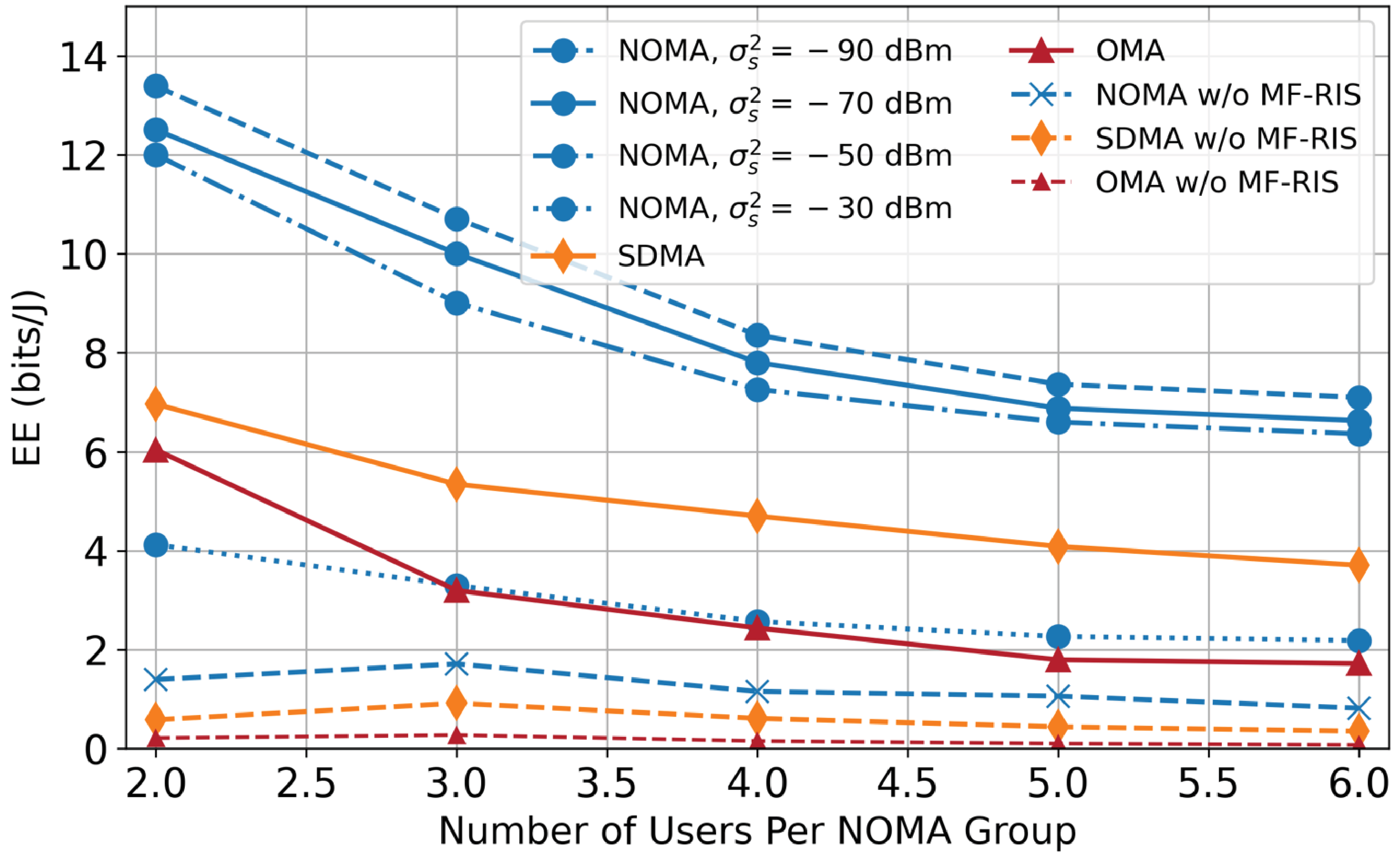} \label{Fig.6}}	
	\caption{(a) Convergence. (b) EE with different strategies (c) EE with different MF-RIS cases. (d) Comparison of NOMA, SDMA, and OMA.}
\end{figure}

	In simulations, we evaluate PMHRL in multi-MF-RISs-assisted downlink NOMA. We consider the BS positioned at $\mathbf{w}_b = [0, 0, 5]$ m, serving $J_k = 2$ users in $K=2$ directions. Then the users are randomly distributed within a circular area of radius 2 m, centered at $[0, 30, 0]$, $[0, 35, 0]$, $[10, 40, 0]$, and $[10, 45, 0]$ m, respectively. The MF-RISs are equipped with $M = 32$ elements, and the BS has $N=6$ antennas. The remaining parameters related to networks are listed in Table \ref{param}. As for PMHRL, learning rates of PPO actor/critic networks are $l^p_{\text{ac}} = 10^{-3}$ and $ l^p_{\text{cr}} = 10^{-4}$, respectively, whereas that of DQN is $l^d=10^{-3}$. The discount factor for both modules is $\gamma^d=\gamma^p = 0.99$. The decay and soft update parameters of DQN are set to $\epsilon_d=10^4$ and $\tau_{\phi}=10^{-2}$, respectively. The experience replay buffer of DQN stores up to $10^6$ samples. The mini-batch mechanism is adopted during training, with a batch size of $64$. $\epsilon_{\max}=1$ and $\epsilon_{\min}=0$. Moreover, we set the clipping ratio to $\Lambda=0.2$, the GAE parameter is $\lambda^p=0.97$, and trajectory length is $10^3$. The weights of each penalty in reward $r(t)$ are $\rho_1=10^{-3}$, $\rho_2=\rho_3=10^{-5}$. Table \ref{EHcoeff} shows the EH efficiency with the corresponding constants of circuit sensitivity $\varpi_1$ and current leakage $\varpi_2$. Higher sensitivity of $\varpi_1$ and higher leakage of $\varpi_2$ result in a lower EH efficiency value. 
	
	Fig. \ref{Fig.4} shows that PMHRL not only achieves a faster convergence but outperforms other methods with the highest EE. Note that under standard Markov assumptions of the bounded state/action spaces and sufficient exploration, the learning process remains well-defined and empirically stable. We can observe that MA-HDRL without parametrized sharing converges more slowly than the other algorithms. This is attributed to the decentralized learning without information sharing, making it challenging to capture effective policies. Also, PMHRL achieves up to a 30$\%$ improvement in EE compared to hybrid DRL due to limited computation and storage capability for tackling high-dimensional state-actions. Moreover, pure PPO architecture \cite{ref21} exhibits the second-lowest EE due to the lack of hybrid learning mechanisms. DQN shows the lowest EE performance due to its large discrete state-action space and quantization errors. Simulations are performed using an Intel Core i7-13700 central processing unit (CPU) and NVIDIA GeForce RTX 4060 graphics processing unit (GPU). The DQN, PPO, Hybrid-DRL, MA-HDRL, and PMHRL possess the computational complexity orders of
$\mathcal{O}(Q N_B N_H (N_L N_H + Q J + L_Q (J+NK+2QMK)))$,
$\mathcal{O}(2 Q N_B N_H (N_L N_H + Q J + NK + 2QMK ))$,
$\mathcal{O}(3 Q N_B N_L N_H^2 + 3 N_H Q J + N_H (J+NK+2QMK))$,
$\mathcal{O}(3 N_B N_L N_H^2 + 3 N_H J + N_H (J+NK+2MK))$, and
$\mathcal{O}(3 N_B N_L N_H^2 + 3 N_H J + 4 N_H (J+NK+2MK))$, respectively. $N_B/N_L/N_L$ are the number of training batch sizes, neural layers, and hidden neurons per layer, respectively. $L_Q$ is the quantization level for continuous variables in DQN. We further quantify the storage memory and execution time per step for each agent across benchmarks. Specifically, the storage memory requirements for DQN, PPO, Hybrid-DRL, MA-HDRL and PMHRL are $\{0.90, 0.39, 0.54, 0.44, 0.45\}$ MB, whereas their respective execution time per step is $\{0.498, 0.0098, 0.019, 0.017, 0.017\}$ s.

    Fig. \ref{Fig.4-1} shows that EE escalates with the increasing numbers of antennas thanks to improved beamforming capability, reaching a peak at $N=6$ before declining as the power consumption begins to outweigh beamforming gains. In addition, we compare the fully-optimized case to the cases without either EH ratio $\alpha_{q,m}$, amplification $\beta_{q,m}^k$, phase-shift $\theta^k_{q,m}$, or deployment $\mathbf{w}_{q}$. Note that "w/o" indicates random decisions. It is evident that full optimization yields the highest EE. In contrast, omitting the optimization of specific parameters leads to noticeable EE degradation. The configuration without optimizing EH ratio results in the lowest EE, as random S-/H-mode leads to inefficient EH for compensating high power consumption. Moreover, different channel conditions of $\beta_0\in\{3,6,12\}$ dB is compared. We can observe that higher $\beta_0$ accomplishes higher EE values thanks to better channel conditions with more LoS channel components.

    Fig. \ref{Fig.5} compares EE of $Q\in\{2,3\}$ MF-RISs under different cases: (1) Optimized case; (2) No EH ($\alpha_{q,m}^k = \alpha_{q,m} = 1, \forall k\in \mathcal{K}$); (3) No amplification ($\beta_{q,m}^k = \beta_{q,m} = 1, \forall k\in \mathcal{K}$); (4) Only reflection capability. The results show that increasing MF-RIS elements initially enhances EE. However, deploying more elements induces higher energy consumption, leading to a degraded EE owing to insufficient support from harvested energy. Notably, the case of $Q=3$ MF-RISs outperforms that of $Q=2$ MF-RISs, attributed to the enhanced spatial diversity and EH gain from multi-MF-RISs. Specifically, when signal amplification is disabled, the reduced signal gain leads to a lower EE than that of the fully-optimized case. Moreover, the signal with only reflection cannot be delivered to users beyond the other side, leading to compellingly huge EE reduction. When EH is fully disabled, it cannot provide extra power, resulting in the lowest EE across all cases.

    Fig. \ref{Fig.6} reveals the EE performance under varying numbers of users per NOMA group and amplification noise levels $\sigma_s^2\in\{-30, -50, -70, -90\}$ dBm. We compare NOMA to the existing multiple access mechanisms of OMA and spatial division multiple access (SDMA) \cite{ref22} with or without deployment of MF-RISs. EE decreases with more users per NOMA group due to insufficient power resources and severe inter-user interference. Moreover, NOMA with MF-RIS deployment achieves the highest EE across all numbers of users, benefiting from its superior spectrum utilization and EH capabilities offered by MF-RISs to SDMA and OMA. In contrast, their counterparts without deploying MF-RIS show significantly lower EE performance. Additionally, larger noise power $\sigma_s^2$ leads to lower EE. Particularly, the amplified noise power of $\sigma_s^2=-30$ dBm induced by MF-RIS results in the lowest EE and asymptotic performance to that of OMA with MF-RIS.

\section{Conclusions}
	We propose a multi-MF-RISs-assisted downlink NOMA networks. An EE optimization problem is formulated, jointly optimizing power allocation, BS beamforming, and MF-RIS configuration of amplification, phase-shift, and EH ratios, as well as positions. To address the high-dimensional and dynamic nature of the complex problem, we have design a PMHRL scheme. Combining both features of PPO and of DQN to respectively handle continuous and discrete action spaces, parametrized sharing is designed to facilitate information exchange between them. Additionally, multi-agent system is leveraged for reducing the overhead. Simulation results validate the superiority of PMHRL outperforming the centralized learning of DQN, PPO, and conventional hybrid DRL in terms of the highest EE. Moreover, the proposed architecture of multi-MF-RISs demonstrates the best performance across various scenarios, including cases without EH, conventional reflective-only RISs, non-amplified signals, and baselines without either RIS or MF-RIS under different multiple access.

\bibliographystyle{IEEEtran}
\bibliography{IEEEabrv}
\end{document}